# Resilience as a Dynamical Property of Risk Trajectories in CPSoS

Elisabeth Vogel
Chair of Wireless Systems
BTU Cottbus-Senftenberg
03046 Cottbus, Germany
Elisabeth.Vogel@b-tu.de

Peter Langendörfer
Chair of Wireless Systems
BTU Cottbus-Senftenberg
03046 Cottbus, Germany
Peter.Langendoerfer@b-tu.de

*Abstract*— Resilience in cyber-physical systems of systems (CPSoS) is often assessed using static indices or point-in-time metrics that do not adequately account for the temporal evolution of risk following a disruption. This paper formalizes resilience as a functional of the risk trajectory by modelling risk as a dynamic state variable. It is analytically shown that key resilience properties are structurally determined by maximum deviation (peak) and effective damping, and that cumulative risk exposure depends on their ratio. A simplified energy-dependent system illustrates the resulting differences in peak magnitude, recovery dynamics, and cumulative impact. The proposed approach links resilience assessment to stability properties of dynamic systems and provides a system-theoretically consistent foundation for the analysis of time-dependent resilience in CPSoS.

## I. Introduction

Resilience is a key concept in safety-critical cyber-physical systems of systems (CPSoS), which are continuously exposed to external disturbances and must limit their effects, recover from impairments, and maintain functionality. Existing resilience concepts typically describe this ability in functional terms such as anticipation, resistance, recovery, and adaptation [1], [2], [3]. However, quantitative operationalizations are often based on static indices, aggregated metrics, or snapshot-based state assessments. The explicit temporal evolution of risk following a disruption is rarely formalized systematically.

CPSoS are inherently dynamic systems whose behaviour is determined not only by individual states but by their state response to external influences. If resilience is a system property, it must therefore be reflected in the temporal development of this response.

This paper formalizes resilience as a functional of the risk trajectory. Instead of evaluating isolated time points, the complete dynamic evolution of risk after a disruption is considered and analysed.

The main contributions are:

- a system-theoretically consistent formalization of resilience as a functional of a dynamic risk trajectory,
- an analytical derivation of structural resilience properties in terms of maximum deviation (peak), effective damping, and cumulative impact,
- an illustrative demonstration using a simplified dynamic system.

This perspective enables a system-theoretically consistent and analytically grounded quantification of time-dependent resilience.

## II. Related Work and Identified Gap

In existing literature, resilience is predominantly conceptualised in functional terms, for example as the ability to anticipate, resist, recover, and adapt to disruptions [2], [4]. In the context of cyber-physical systems and critical infrastructures, such concepts are frequently applied for structural or organisational assessments. Numerous studies propose quantitative indicators or aggregated metrics that incorporate different system dimensions [5], [6].

Despite these developments, many quantitative approaches rely on static indices, weighted score models, or snapshot-based assessments of individual system states. The explicit temporal evolution of risk following a disruption is often treated implicitly or evaluated through simulation-based analyses. Although some studies investigate curve progressions or performance degradation models, these representations are rarely embedded in a consistent dynamic system framework [5], [6]. As a result, the structural relationship between resilience indicators and the underlying system dynamics often remains unclear.

CPSoS are inherently dynamic systems whose behaviour is characterised by their state response to external influences. If resilience is understood as a system property, it must therefore be reflected in the temporal development of this response. Despite numerous quantitative approaches, a system-theoretically consistent formalisation of resilience as a functional of a dynamic risk trajectory remains insufficiently developed. This methodological gap forms the starting point for the formal modelling presented in the following section.

## III. Risk as a Dynamical Variable

CPSoS can be described as dynamic systems whose behaviour is determined by time-dependent state changes. In general form, such a system can be represented as

This work was supported by the Federal Ministry of Transport (BMV) under research grant number 01FV2063A.

$$\frac{dx}{dt} = f(x(t), d(t)), \qquad (1)$$

where $x(t)$ denotes the system state vector and $d(t)$ represents an exogenous disturbance.

The behaviour of a dynamic system is characterised by its state response to external inputs. For safety-related or safety-critical systems, risk can be modelled as a time-dependent state variable. Accordingly, a risk dynamic of the form

$$\frac{dr}{dt} = g(r(t), d(t)), \qquad (2)$$

is assumed, where $r(t)$ describes the time-dependent risk level.

While $x(t)$ represents the complete system state, $r(t)$ denotes a reduced, risk-relevant state variable derived from the underlying system dynamics.

After a disturbance occurs at $t_0$, the resulting state response

$$r(t), t \geq t_0, \qquad (3)$$

is referred to as the risk trajectory. This trajectory describes the complete temporal evolution of risk following a disruption.

An isolated consideration of individual time points is insufficient to characterise resilience. Two systems may reach the same final state but exhibit significantly different maximum deviations, recovery dynamics, or cumulative stress profiles. Resilience is therefore not a point value but a property of the entire trajectory.

Formally, resilience can be represented as a functional over the risk trajectory:

$$Res = F(r(t)). \qquad (4)$$

The specific form of this functional depends on structural properties of the underlying system dynamics. In the following section, key structural quantities of the risk trajectory are derived analytically.

## IV. Structural Derivation of Time-Dependent Resilience Metrics

Based on the description of resilience as a functional of the risk trajectory introduced in Section III, the structural properties of this trajectory are analysed analytically. The objective is to derive measurable quantities that characterise resistance, recovery dynamics, and cumulative impact.

To this end, the maximum deviation following a disturbance is first considered, followed by the recovery dynamics and finally the integrated risk exposure.

### A. Resistance: Peak Deviation

After a disturbance occurs at t_0, the risk trajectory deviates from its equilibrium to a maximum extent. This maximum deviation is defined as

$$r_0 = \max_{t \geq t_0} r(t). \qquad (5)$$

The value $r_0$ characterises the immediate vulnerability of the system to the disturbance. A small $r_0$ indicates high resistance, whereas a large $r_0$ reflects a severe initial impairment.

Structurally, the peak depends on the sensitivity of the system to external influences as well as on existing buffering or anticipatory mechanisms. However, the peak alone does not allow conclusions about the duration or cumulative effect of the disturbance. A complete characterisation of resilience therefore also requires consideration of the return dynamics to equilibrium.

### B. Recovery and Effective Damping

After the initial deviation, the return to equilibrium is considered. The risk dynamics can generally be described as

$$\frac{dr}{dt} = g(r(t)), \qquad (6)$$

assuming that a stable equilibrium exists at r=0.

In the vicinity of a stable equilibrium, nonlinear systems can be locally approximated by their linearisation, leading to exponential decay behaviour determined by the dominant eigenvalue [7]. For small deviations from equilibrium, the dynamics can therefore be approximated by

$$\frac{dr}{dt} = -\lambda r, \qquad (7)$$

where $\lambda > 0$ characterises the local stability property of the system. This representation corresponds to the standard linear approximation of stable dynamic systems near an equilibrium point.

The solution of this equation is

$$r(t) = r_0 e^{-\lambda t}. \qquad (8)$$

The parameter $\lambda$ can be interpreted as an effective damping coefficient. A large value of $\lambda$ results in a rapid return to the safe operating region, whereas a small value of $\lambda$ implies prolonged exposure to elevated risk.

This shows that resilience cannot be characterised by an isolated time point but is structurally determined by the local stability properties of the system dynamics.

### C. Cumulative Impact

To evaluate the total effect of a disturbance, the cumulative risk exposure is defined as

$$Impact = \int_0^\infty r(t) dt. \qquad (9)$$

This quantity captures the overall impact of the disturbance over time and accounts for both the maximum deviation and the duration of the recovery phase.

Using the local approximation derived in Section 4.B,

$$r(t) = r_0 e^{-\lambda t}, \qquad (10)$$

we obtain

$$\text{Impact} = \int_0^\infty r_0 e^{-\lambda t} dt = \frac{r_0}{\lambda}. \tag{10}$$

This expression shows that the cumulative impact depends structurally on two quantities: the maximum deviation $r_0$ and the effective damping $\lambda$.

A system with a small peak but weak damping may produce a cumulative impact comparable to that of a system with a larger peak but faster recovery. Resilience therefore emerges from the interaction between peak deviation and damping.

Nonlinear effects or finite time horizons may alter the exact quantitative value. However, the structural dependence on peak deviation and damping remains.

### D. Structural Interpretation and Functional Perspective

The preceding derivations show that resilience is structurally determined by maximum deviation and local stability properties. Anticipatory mechanisms primarily reduce the peak deviation $r_0$, while adaptive or feedback structures increase the effective damping $\lambda$.

Resilience can therefore be interpreted as a structural property of the underlying system dynamics, directly linked to local stability characteristics.

The derived quantities represent concrete manifestations of the functional introduced in Section III,

$$\text{Res} = F(r(t)). \tag{11}$$

While the peak captures the maximum deviation and the damping reflects local stability, the integral measure represents the combined structural effect of both properties. Resilience can thus be understood as a functional over the risk trajectory, whose specific form is determined by structural system parameters.

To illustrate these structural relationships, the following section introduces a simplified system model that exemplifies the interaction of peak deviation, effective damping, and cumulative impact.

## V. ILLUSTRATIVE SCENARIO: ENERGY-CONSTRAINED SYSTEM

To illustrate the structural relationships derived in Section IV, a simplified energy-dependent system is considered. The aim is to demonstrate how different structural characteristics influence maximum risk deviation, recovery dynamics, and cumulative impact.

The system comprises a limited energy storage whose state $E(t)$ is determined by consumption and exogenous energy input. The energy input is modelled as periodic solar feed-in.

A decrease in the energy level leads to an increase in systemic risk. Risk is modelled as a normalized state variable $r(t)$ that increases as available energy decreases. The system is exposed to external disturbances and, depending on the scenario, is equipped with different anticipatory and adaptive mechanisms.

In the following, three structurally distinct system configurations are examined to analyse the influence of these mechanisms on peak deviation, effective damping, and cumulative risk exposure.

### A. System Model

The system under consideration comprises a limited energy storage with energy state $E(t)$. The energy dynamics are described in simplified form as

$$\frac{dE}{dt} = P_{in}(t) - P_{load}(t), \tag{12}$$

where $P_{in}(t)$ denotes the exogenous energy supply (modelled here as periodic solar feed-in) and $P_{load}(t)$ represents the system's energy consumption.

A decrease in the energy level results in an increase in systemic risk. Risk is modelled as a normalized state variable $r(t)$ that increases monotonically as available energy decreases. Formally, this relationship can be expressed as

$$r(t) = \phi(E(t)), \tag{13}$$

where $\phi(\cdot)$ is a monotonically decreasing function.

The system is subjected to a continuous load and reaches an energetic equilibrium in steady-state operation. Consequently, the risk trajectory does not necessarily converge to zero but to an operating point determined by the balance between energy supply and consumption.

Different structural characteristics of the system, particularly with respect to anticipation and adaptivity, influence both the maximum risk deviation and the recovery dynamics following energy disturbances.

### B. Scenario Definition

For the analysis, three structurally distinct system configurations are considered, differing in their anticipatory and adaptive mechanisms.

#### Case 1 – Passive Configuration

In the first configuration, the system does not employ active stabilisation mechanisms. Energy consumption remains constant, regardless of the current energy level. Consequently, external fluctuations in energy supply directly result in significant energy depletion.

If critical energy thresholds are reached, risk increases sharply. Neither anticipatory nor adaptive measures provide damping of the system response.

#### Case 2 – Reactive Configuration

In the second configuration, the system incorporates reactive adaptation mechanisms. When the energy level decreases, non-critical subsystems are deactivated, thereby reducing energy consumption.

These measures exert a stabilising influence but only take effect after the energy level has already declined substantially. As a result, the peak deviation is moderately reduced, and recovery dynamics improve compared to Case 1.

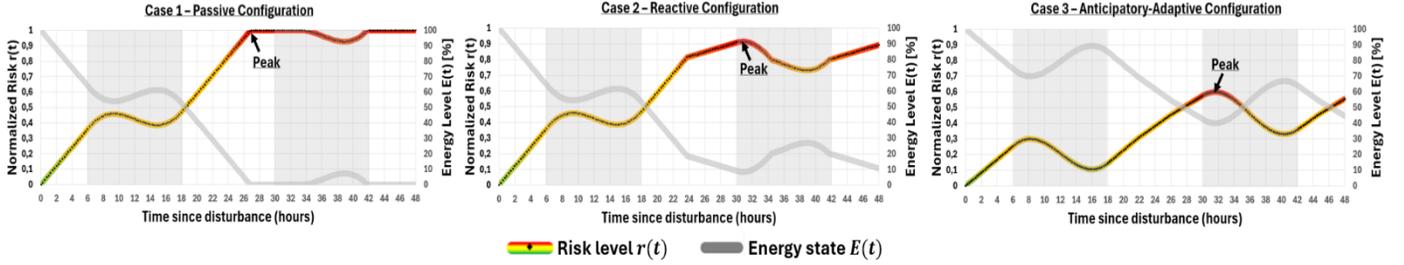

Fig. 1: Risk and energy trajectories for three structural system configurations: (Case 1) passive, (Case 2) reactive, and (Case 3) anticipatory-adaptive. The plots illustrate differences in peak deviation and recovery dynamics following energy disturbances. The periodic solar input is indicated by shaded regions. While the passive configuration exhibits high peak risk and prolonged exposure, the anticipatory-adaptive system stabilizes within a bounded operating regime. The observed trajectories reflect the structural dependency of cumulative impact on peak deviation and effective damping.

Case 3 – Anticipatory-Adaptive Configuration

In the third configuration, adaptive measures are combined with anticipatory strategies. Energy consumption is adjusted at an early stage, for instance through prioritisation of critical functions or proactive utilisation of available energy resources.

This configuration both limits the maximum risk deviation and increases the effective damping. The system stabilises within a bounded steady-state operating regime characterised by limited risk oscillations.

### C. Trajectory-Based Analysis

The three system configurations lead to significantly different risk trajectories following energy disturbances. The corresponding energy and risk trajectories are shown in Fig. 1.

In Case 1, the absence of stabilisation results in a pronounced risk deviation. The peak $r_0$ is high and recovery dynamics are weak. Effective damping is low, leading to prolonged risk exposure and a high cumulative impact.

In Case 2, reactive adjustment mechanisms reduce the peak deviation and improve recovery dynamics compared to Case 1. However, stabilisation occurs only after a substantial energy decline, so that both the maximum deviation and the duration of exposure remain significant. The cumulative impact decreases relative to Case 1 but remains elevated.

In Case 3, the combination of anticipatory and adaptive mechanisms significantly limits the maximum deviation. Early adjustment increases the effective damping, resulting in faster recovery. The risk trajectory stabilises within a bounded operating regime characterised by limited oscillations. Both the peak deviation and exposure duration are reduced, yielding the lowest cumulative impact among the three configurations.

The analysis demonstrates that resilience is not determined by a single property but by the interaction of maximum deviation $r_0$ and effective damping $\lambda$. Systems with similar peaks may exhibit substantially different cumulative impacts due to differences in damping. The observed trajectories therefore illustrate the structural dependence of cumulative impact on peak deviation $r_0$ and effective damping $\lambda$, derived in Section IV and expressed as

$$Impact = \frac{r_0}{\lambda}. \qquad (14)$$

The trajectories thus provide an intuitive confirmation of the trajectory-based resilience perspective.

## VI. DISCUSSION

This paper proposes a trajectory-based perspective on resilience, in which resilience is understood not as a static index but as a structural property of the temporal state response. In contrast to point-based assessment approaches, considering the complete risk trajectory enables a differentiated analysis of maximum deviation, recovery dynamics, and cumulative impact. This perspective shifts the focus from isolated performance indicators to the underlying system dynamics that generate observable resilience behaviour.

The derived dependence of cumulative impact on peak deviation and effective damping is based on a local stability approximation. Although real systems are generally nonlinear and may exhibit higher-order effects, the structural relationship between maximum deviation and recovery rate remains conceptually valid. The analysis highlights that resilience cannot be reduced to a single scalar metric without considering its dynamic origin. Instead, resilience emerges from the interaction between resistance to disturbance and the system's ability to restore stability.

The proposed representation should therefore be understood as an analytical framework rather than a complete representation of all system details. It provides structural insight into how anticipatory and adaptive mechanisms influence resilience properties through their effect on peak deviation and effective damping. In this sense, the framework establishes a direct conceptual link between resilience assessment and classical stability theory.

The presented energy scenario serves to illustrate these structural relationships and does not constitute empirical validation. The simplified model abstracts from domain-specific complexities in order to make the structural dependencies transparent. Future work may extend the trajectory-based perspective to higher-dimensional state spaces, incorporate stochastic disturbances, and integrate data-driven methods for estimating the relevant parameters. In particular, the combination of the analytical framework with learning-based approaches may enable the practical quantification of time-dependent resilience in real-world CPSoS.

## VII. Conclusion

This work establishes a trajectory-based formalization of resilience as a property of temporal risk dynamics in cyber-physical systems of systems. Instead of relying on point-based assessment approaches, resilience is formulated as a functional of the state response to disturbances, thereby embedding resilience explicitly within a dynamic system framework.

The analytical derivation demonstrates that key resilience properties arise structurally from maximum deviation and effective damping. This formulation reveals that cumulative impact is not an independent metric but emerges from the interaction between resistance and recovery characteristics. In doing so, the framework provides a transparent structural interpretation of resilience and explicitly links it to local stability properties of dynamic systems.

By grounding resilience assessment in system dynamics, the proposed approach offers a conceptually coherent and analytically tractable basis for the quantitative evaluation of time-dependent resilience. The trajectory-based perspective thus contributes to a more systematic integration of resilience analysis and stability theory in complex CPSoS.

Future research may extend the proposed trajectory-based formulation to higher-dimensional state spaces and more complex system representations. In particular, the integration of stochastic disturbances and nonlinear system dynamics could provide a more detailed characterisation of resilience behaviour in real-world CPSoS. Furthermore, data-driven approaches may support the empirical estimation of key parameters such as peak deviation and effective damping, enabling the practical application of the proposed metrics in operational resilience assessment.